\documentclass[runningheads]{llncs}
\usepackage{xcolor}
\definecolor{myblue}{RGB}{33,113,181} 
\usepackage{tikz}
\usetikzlibrary{arrows.meta, positioning}

\usepackage{changepage}
\usepackage{tabularx}
\usepackage[T1]{fontenc}
\usepackage{comment}
\definecolor{myblue}{RGB}{33,113,181} 
\usepackage{tikz}
\usetikzlibrary{arrows.meta, positioning}
\usepackage{booktabs}
\usepackage{xurl}
\usepackage{hyperref}
\usepackage{microtype}
\usepackage{graphicx}
\usepackage{subcaption}
\usepackage{float}

\usepackage{xurl}
\begin{document}
\title{PAGURI: A User Experience Study of Creative Interaction with Text-to-Music Models}
%
%
\author{Francesca Ronchini\orcidID{0000-0001-6897-1645} \and
Luca Comanducci\orcidID{0000-0002-4167-5173} \and
Gabriele Perego\and
Fabio Antonacci\orcidID{0000-0003-4545-0315}}
\authorrunning{Francesca Ronchini et al.}
%
\institute{Dipartimento di Elettronica, Informazione e Bioingegneria (DEIB), Politecnico di Milano
Piazza Leonardo Da Vinci 32, 20133 Milan, Italy\\\email{\{name.surname\}@polimi.it}}

\maketitle              
\begin{abstract}
In recent years, text-to-music models have been the biggest breakthrough in automatic music generation. While they are unquestionably a showcase of technological progress, it is not clear yet how they can be realistically integrated into the artistic practice of musicians and music practitioners. This paper aims to address this question via Prompt Audio Generation User Research Investigation (PAGURI), a user experience study where we leverage recent text-to-music developments to study how musicians and practitioners interact with these systems, evaluating their satisfaction levels. We developed an online tool through which users can generate music samples and/or apply recently proposed personalization techniques based on fine-tuning to allow the text-to-music model to generate sounds closer to their needs and preferences. Using semi-structured interviews, we analyzed different aspects related to how participants interacted with the proposed tool to understand the current effectiveness and limitations of text-to-music models in enhancing users' creativity. 
Our research centers on user experiences to uncover insights that can guide the future development of TTM models and their role in AI-driven music creation. Additionally, they offered insightful perspectives on potential system improvements and their integration into their music practices. The results 
 obtained through the study reveal the pros and cons of the use of TTMs for creative endeavors. Participants recognized the system’s creative potential and appreciated the usefulness of its personalization features. However, they also identified several challenges that must be addressed before TTMs are ready for real-world music creation, particularly issues of prompt ambiguity, limited controllability, and integration into existing workflows.

\end{abstract}

\section{Introduction 
}
\label{sec:intro}

Since the 1950s, computer-based music generation has been an interest of both the music and computer science research communities~\cite{hiller1957musical,mathews1969technology}. Deep learning has been widely applied to music generation, rapidly becoming a well-researched branch of the field~\cite{briot2020deep}.
In this context, the latest breakthrough has been the recent introduction of Text-To-Music (TTM) models, which enable the generation of raw audio musical signals based on input text prompts describing the desired music. 

The introduction of TTM models has reduced the technical skills required to use music generative models, making it necessary, for the first time, to consider whether and how AI can be integrated into music creation practices.
Although generative models for music are gaining more and more popularity, there is still a lack of comprehensive research on this topic. We advocate that it is important to conduct this type of research closely with potential users of TTM models. In this study, we define a user as a person with different levels of musical expertise who interacts with the considered TTM model. This approach helps develop tools that not only showcase the impressive capabilities of the technology in the music field but also highlight how these tools can be viewed and perceived as musical instruments for creating music.

In this paper, we introduce the Prompt Audio Generation User Research Investigation (PAGURI). Through PAGURI, our goal is to analyze how users engage with TTM models, their perceptions of the tools, and their potential integration into the music creation process based on user feedback. While TTM model creators primarily focus on technical performance~\cite{kamath2024sound,zang2024interpretation}, we believe it is equally important to understand user interactions, assess the benefits and limitations of current state-of-the-art models, and evaluate whether the existing metrics are adequate~\cite{tailleur2024correlation,gui2024adapting}. 
A deeper understanding of these interactions will enable more informed, human-centered model development, moving beyond audio quality to consider real-world usability and the interpretation gap in TTM generative models~\cite{zang2024interpretation}. Our study also focuses on the potential benefits of adding features such as personalization techniques, which involve providing audio input alongside the prompt. We advocate that the current focus on using a text prompt as the only input for TTM models is too limiting~\cite{nistal2024diff,zang2024interpretation}, and understanding how such features affect user experience is critical for designing systems that are more adaptable and engaging. By focusing our research on user experiences, we aim to provide interesting insights that can contribute to the future evolution of TTM models and their role in AI-driven music creation. To the best of our knowledge, no prior study has specifically addressed the interactions between users and TTM models, making our investigation a novel contribution to the field. This study is focused on the raw audio domain, while symbolic music will be considered in future work.

{More specifically, the research gap and research questions that this paper tries to fill and answer, respectively, are the following: \textit{Are 
 text-to-music models effectively usable for creative music endeavors?} \textit{Is the possibility of personalizing the generative models with audio selected by the user efficient from the point of view of creativity?}\textit{What insights can we gather from analyzing text-to-music models by focusing on user experience studies instead of only on quantitative metrics?}}
For this purpose, we developed an interface that allows users to generate audio samples by specifying their desired sound through text prompts. Additionally, the user can upload up to $5$ desired audio samples to personalize the TTM generative model.
For the study, we use AudioLDM2~\cite{liu2023audioldm2} as a generative model and a TTM personalization technique proposed in ~\cite{plitsis2023investigating} to let the users fine-tune the model according to the music samples of their choice. 
We then conducted a user experience study 
where users could interact with PAGURI, and we analyzed their experience using semi-structured interviews~\cite{kamath2024sound,sarmento2024between,hashim2023music,tchemeube2023evaluating}. 
Specifically, we first quantify their background related to music and AI tools, and then we analyze their level of satisfaction after each use of the TTM model. 
Finally, we let them answer a questionnaire analyzing the whole experience and gathering information related to the perceived usability of the TTM model in music practice.
The user experience study was conducted both online and in person, following the same procedure in both formats. The call for participation was distributed through email lists relevant to the field, aiming to reach a diverse and varied group of participants for the user experience study. The participant pool consisted of 24 individuals, primarily Italian, with most being students enrolled in the M.Sc. program in Music and Acoustic Engineering at the Politecnico di Milano, Milan, Italy. The time-consuming nature of the experiment, conducted live even for the online sessions, made it challenging to reach a broader audience. While we acknowledge that this may limit the generalizability of the results, we believe that the participants’ unique backgrounds in both music and technology closely reflect the characteristics of the target audience for text-to-music models.  {Nevertheless, we believe that the number of participants is sufficient for an exploratory analysis of the problem of interaction with TTM models.}
We consider the findings of this study a valuable contribution to the research community, advancing the understanding of user experience in generative models and their role in AI-driven music creation.
Additionally, they represent an initial step in fostering practical discussions regarding user interactions with such models.

We can then summarize the objectives, methodologies, and contributions of our manuscript as follows:

\begin{itemize}

 \item \textbf{Objectives
}: to investigate the creative interaction of musicians and music practitioners with a TTM system, focusing on prompt consistency, personalization, and \mbox{system integration.}
 \item \textbf{Methodology}: a mixed-method user study with 24 participants, combining Likert-scale ratings, open interviews, and thematic analysis.
 \item \textbf{Contributions}: empirical insights into user expectations and challenges with TTM models, an analysis of model personalization as a creative instrument, and recommendations for future human–AI co-creative system design.
\end{itemize}

\textls[-15]{Code and Supplementary Material containing the answers to all questions, feedback, and further details are available on the accompanying website 
\textls[-15]{{{\url{https://ronfrancesca.github.io/PAGURI/} (accessed on 25 August 2025)}}}. {The rest of the paper is organized as follows. In Section~\ref{sec:back}, we present the necessary background related to text-to-music models, personalization techniques, and human–AI interaction studies in order to make the manuscript as self-contained as possible. In Section~\ref{sec:method}, we present the methodology used to design and conduct the experimental study, while in Section~\ref{sec:results}, we describe the results obtained. Section~\ref{sec:discussion} provides the reader with a discussion relative to the results obtained by the experiments; finally, Section~\ref{sec:concl} draws some conclusions.}}

\section{Background}
\label{sec:back}

In this section, we provide the reader with the necessary background related to text-based generative music models and techniques used to personalize them.

\subsection{Text-to-Music Models}
\label{subsec:TTM}

Recently, a large number of multimodal models generating music conditioned on textual descriptions have been proposed. While describing all the details of these models is out of the scope of this paper, we provide a brief overview. TTM models can be decomposed in two parts~\cite{afchar2025ai}: an audio encoder that compresses the raw signal into a quantized representation and a separate model that learns to generate such representations conditioned on text. TTM techniques differ depending on the audio quantizer used, normally based on Residual Vector Quantization~\cite{barnes1996advances}, such as Encodec~\cite{defossez2023high}, Soundstream~\cite{zeghidour2021soundstream}, or DAC~\cite{kumar2023highfidelity}. Also, an important distinction concerns the model used to generate the audio tokens, which could either be a transformer-based architecture~\cite{vaswani2017attention} or a generative model, such as a diffusion-based one~\cite{ho2020denoising}.

Many works have applied such techniques to the task of raw audio music and audio synthesis, employing different types of architectures. The first model proposed was AudioLM~\cite{borsos2023audiolm}, a multi-stage transformer-based model operating on tokens extracted from audio via SoundStream~\cite{zeghidour2021soundstream} and from text via w2v-BERT~\cite{chung2021w2v}. This methodology was further developed in MusicLM~\cite{agostinelli2023musiclm}, where the joint music-text embedding model MuLan~\cite{huang2022mulan} was applied to overcome data scarcity. Similar transformer-based approaches were proposed for general audio synthesis using auto-regressive AudioGen~\cite{kreuk2022audiogen} and 
MusicGen~\cite{copet2024simple}, as well as non-auto-regressive models such as MAGNeT~\cite{ziv2024masked}.
Several diffusion-based models were also proposed. DiffSound~\cite{yang2023diffsound}, the first one, considered audio generation by proposing a text-conditioned diffusion model operating on tokenized audio. Later on, several diffusion-based approaches were also proposed for text-to-music generation, such as Make-an-Audio~\cite{huang2023make}, AudioLDM~\cite{liu2023audioldm}, and their evolutions Make-an-Audio 2~\cite{huang2023make2} and AudioLDM2~\cite{liu2023audioldm2}, MUSTANGO~\cite{melechovsky-etal-2024-mustango}, and Stable Audio Open~\cite{evans2024stable}. More recently, several commercial solutions were also proposed, such as Suno AI~\cite{suno2024}, which raised USD 125 million in funding, and Udio~\cite{udio2024}. 

\subsection{Personalization Techniques}
\label{susec:perso}

Personalizing a model involves adapting pre-trained generative models to new examples that were not part of the original training dataset. In the context of language-based generative models, the task was first introduced for images in~\cite{gal2023an}, where the authors proposed the \textit{Textual-Inversion} technique. This technique allows for injecting into the model information from data that it did not encounter during training. This is performed by using a few images whose text-embedding representations, or \textit{pseudo\-words} (i.e., tokens), are learned in the embedding space of the text encoder. This technique is limited by the expressive capabilities of the pre-trained generative model. To overcome this limitation, the \textit{DreamBooth} approach was proposed in~\cite{ruiz2023dreambooth}. In this case, the desired examples are represented using rare token identifiers, and, consequently, the considered generative model (usually diffusion-based) is fine-tuned to learn the desired content.

Later on, personalization techniques were further developed for TTM generative models. This advancement was initially showcased in~\cite{plitsis2023investigating},  where Textual Inversion and DreamBooth techniques were integrated with AudioLDM~\cite{liu2023audioldm}. The objective was to enhance the model's ability to learn new music samples and conduct style transfer. 

\subsection{Human–AI Interaction and Collaborative Creativity}
Previous research has explored the perception of AI-generated music~\cite{chu2022empirical} and the opportunities and challenges of AI for music creation from a human perspective~\cite{newman2023human}. The interaction between humans and AI in music generative models has also been explored, particularly in the context of creative collaboration. In~\cite{morreale2016collaborating}, a novel generative system is introduced, using emotion-based representations as an interactive metaphor to enable users to shape musical outcomes. In~\cite{louie2020novice}, the authors investigate the needs of novice users when co-creating music with deep generative models, proposing AI-steering tools that facilitate iterative, real-time creative control. {In~\cite{louie2022expressive}, it is discussed how to} evaluate generative models and AI-steering interfaces in the context of downstream creative goals, specifically examining how novice composers express emotions in music through a combination of subjective self-reports and objective listener judgments. {In \cite{zhou2021interactive}, the authors explore} a novel approach to enhance user interaction by allowing direct control over the model’s sampling behavior, refining the creative process. In~\cite{huang2020ai}, an in-depth study on human–AI co-creation is presented, identifying the challenges musicians and developers face when composing with AI, as well as strategies they employ to repurpose AI capabilities, while~\cite{sarmento2024between} investigates listeners' perspectives on AI-versus human-generated progressive metal, providing further insights into human perception of AI-generated music.

While user interaction with text-based generative models has been explored in the image domain~\cite{feng2023promptmagician,brade2023promptify}, where datasets of Text-To-Image (TTI) prompts with corresponding preferences given by real users over AI-generated images~\cite{kirstain2024pick} have also been released, research on text-based models in other domains remains limited. As interpretability challenges in generative models persist~\cite{zang2024interpretation}, it is crucial to address this gap. This study aims to advance the field and offer deeper insights into the extent to which text-based models can \mbox{enhance creativity.}

\subsection{Interfaces for Human–AI Interaction}
\label{subsec:humanAI}

Several interfaces have been proposed to improve interaction between generative models and users~\cite{bougueng2022calliope,simon2008mysong,huang2019human,louie2020novice,rau2022visualization,zhang2021cosmic,Zhou2020generative,zhou2021interactive,yakura2023iteratta} and AI-assisted music composition~\cite{rau2025maico}, alongside creative supporting tools designed to enhance collaborative creativity~\cite{chung2021intersection,frich2019mapping,kamath2024sound}. 
With the growing popularity of generative applications, researchers have increasingly explored these developments by leveraging recent advances in human–AI co-creativity, proposing design principles for generative AI interfaces~\cite{weisz2023toward}. 

Among the proposed studies, the closest work to the one proposed in this paper is IteraTTA~\cite{yakura2023iteratta}, an interface for TTM generation enabling the iterative exploration of prompts and audio priors. Different from PAGURI, IteraTTA did not contain the possibility of personalizing the model using audio samples of choice. Moreover, our goals and design strategy are different. The objective of IteraTTA was mainly to design an interface for supporting novice users, analyzing user behavior ex-post. In our study, the simple PAGURI interface was designed to study how music practitioners interacted with these models. For this reason, the whole user experience test was conducted live using semi-structured interviews, containing both closed and open-ended questions~\cite{kamath2024sound,hashim2023music,sarmento2024between,tchemeube2023evaluating}.

\section{Study Design and Method}
\label{sec:method}

This section presents the study design and the methodology proposed. Section \ref{subsec:reasons} briefly introduced the model and personalization technique selected for this study; in Section \ref{subsec:interface}, we introduce the PAGURI interface, while Section \ref{subsec:procedure} presents the procedure followed to conduct the interactive experiment.
 

\subsection{Model and Personalization Technique}
\label{subsec:reasons}
For this study, we selected AudioLDM2~\cite{liu2023audioldm2} as the generative model and DreamBooth~\cite{ruiz2023dreambooth} as the personalization technique.

AudioLDM2 is capable of generating both general audio and music, demonstrating strong performance in music generation, outperforming existing music generation models at the time of this experiment~\cite{liu2023audioldm2}. Our study specifically focuses on a broad range of users, including musicians, music producers, and sound designers. Given the diverse range of musical styles and sonic textures, AudioLDM2’s versatility in generating both structured and unstructured sounds makes it a valuable tool for various music-related applications.
Additionally, AudioLDM2 enables fine-tuning with personalization techniques, offering a reasonable compromise between computational resources needed and the training time required for fine-tuning.

DreamBooth is a personalization technique for fine-tuning a pretrained text-to-audio model using audio samples as input. Specifically, a unique embedding vector $v^*$ is associated with a placeholder string $S^*$, expanding the text encoder’s parameter space. The model is then optimized by updating the denoising network’s weights $\phi$, allowing it to generate sounds that match the provided characteristics while responding to diverse text prompts. For further details on the personalization technique, the reader is referred to~\cite{ruiz2023dreambooth,plitsis2023investigating}.

\subsection{PAGURI Interface}
\label{subsec:interface}
\begin{figure}[h]
    \centering
\includegraphics[width=\linewidth]{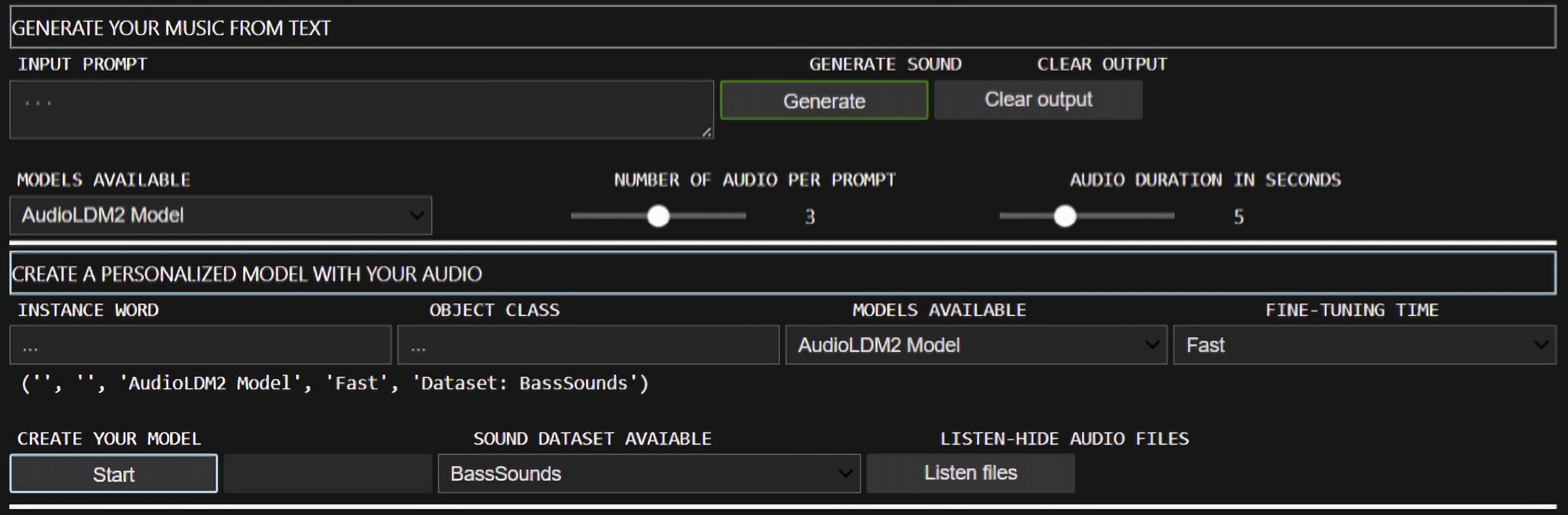}
    \caption{The user interface employed to carry out the PAGURI study. The upper part corresponds to prompt-based music generation using AudioLDM2, while the bottom part corresponds to the personalization of the model via DreamBooth, with the possibility to upload audio samples of choice.}
    \label{fig:paguri_GUI}
\end{figure}

The Prompt Audio Generation User Research Investigation (PAGURI) proposed in this paper presents an interface through which users can generate 
audio samples using state-of-the-art TTM models, depicted in Figure \ref{fig:paguri_GUI}. Through the interface, users can generate audio samples, giving textual input prompts of their choice to the generative model. They can specify the number of audio tracks to generate and the duration for each iteration.
Additionally, through the same interface, users can upload up to $5$ preferred audio files to personalize AudioLDM2. 
Subsequently, they can obtain a personalized model using the DreamBooth personalization technique as implemented  in~\cite{plitsis2023investigating}. In this case, the user can specify an \textit{instance word} used as an identifier to be attached to the new audio files and an \textit{object class}, which indicates a more general description of the audio, which may indicate the instrument, genre, or musical style of the concepts. 
Each user can select to either personalize the off-the-shelf AudioLDM2 model or a previously personalized (by the same user) version of it.
The fine-tuning duration can be chosen between \textit{Fast, Medium}, and \textit{Slow}. Each of them applies the DreamBooth personalization procedure for different durations, ranging from a minimum of $3$ min (\textit{Fast}) to a maximum of $15$ min per iteration (\textit{Slow}). 
Although longer fine-tuning times could improve audio quality, we limited the minimum and maximum durations to 3 and 15 min, respectively, in order to maintain the interactive nature of the study. 
We developed the interface following the AI controllability principles outlined in~\cite{weisz2023toward}. For the scope of this study, we prioritized the principle related to \textit{designing for exploration} by integrating multiple outputs and controls~\cite{weisz2023toward}. The ability to generate multiple outputs allows users to explore different possibilities and better understand the model’s range of capabilities. Providing controls enables interactive engagement, allowing users to navigate the space of possible outputs for a given input. For this aspect, we emphasized \textit{human control} by implementing both generic controls, applicable across different generative technologies, and domain-specific controls~\cite{kamath2024sound}, especially in the case of the personalization technique for the generative audio model. This approach allows for greater flexibility in adapting the model’s responses to specific user preferences and requirements. For generic controls, we took inspiration from models available through platforms like Hugging Face\footnote{\url{https://huggingface.co/spaces/facebook/MusicGen} (accessed on 25 August 2025), \url{https://huggingface.co/spaces/haoheliu/audioldm2-text2audio-text2music} (accessed on 25 August 2025)}, ensuring that users have familiar and intuitive control options similar to those found in these interfaces.

\textls[-15]{The PAGURI interface has been developed using the \textit{Python} 3.9 programming language, with the use of \textit{Jupyter Notebook} 6.5.* and the \textit{ipywidgets} 8.1.x library\footnote{\url{https://github.com/jupyter-widgets/ipywidgets} (accessed on 25 August 2025)}. {The open-source code of the interface is freely available online {\url{https://github.com/RonFrancesca/PAGURI/blob/main/PAGURI_interface.ipynb}}} (accessed on 25 August 2025). 
 A detailed presentation and images showing the PAGURI interface can be found in the Supplementary Material} \url{https://ronfrancesca.github.io/PAGURI/} (accessed on 25 August 2025).

\subsection{Experiment Procedure}
\label{subsec:procedure}

The participant could choose to conduct the user experience either in person or online, scheduling a one-hour slot by agreeing on a day and time with the experimenter.
The in-person experiment was conducted using a laptop in a meeting room at the Department of Electronics, Information, and Bioengineering at Politecnico di Milano, with the participant having full control of the laptop. For the online procedure, participants were given remote control of the laptop's desktop and could interact with the GUI interface independently. {The participants were therefore free to use their preferred setup in terms of browser and/or headphones when performing the experiment.}

The whole procedure 
is divided into three steps: 

\begin{enumerate}

    \item \textbf{Preliminary analysis}: The participants were introduced to the study, and they were required to answer a brief questionnaire containing demographic questions (age, nationality, etc.). To frame the knowledge of participants concerning TTM models, we asked them to answer a second questionnaire regarding their musical knowledge and experiences with AI tools, using questions partially taken from the Goldsmiths Musical Sophistication Index (Gold-MSI)~\cite{mullensiefen2014musicality}. 
    
    \item \textbf{Text-to-music interaction}: 
    Participants were asked to interact with the PAGURI interface. At each iteration, participants had the option to input a new prompt into the TTM model, choose whether or not to personalize the model with their selected audio, and then generate a desired number of new audio samples. 
    After each generation iteration, participants were asked to complete the \textit{Model Evaluation Survey}, available on the Supplementary Material, expressing their satisfaction with the generated audio samples regarding their consistency with the input prompt, audio quality, and alignment with their general expectations. To balance speed with quality, we limited the available audio to 5 samples based on observed interaction times in the \textit{Fast} mode, where the average time between fine-tuning and completing the intermediate questionnaire was 8 min. 
    
    \item \textbf{Final analysis.}
   Upon completion of the experiment, participants were requested to complete a questionnaire regarding their satisfaction with the entire interaction experience with the TTM model via PAGURI. 
   We also asked for open-answer comments and suggestions regarding possible applications and the inclusion of TTM models in artistic practice. 
\end{enumerate}

 The whole procedure lasted approximately one hour. 
 The number of iterations that the participant had with the models during step 2 
 was variable and chosen by the users according to their desire. To analyze repeated trends, we carried out an inductive, open-coding, thematic analysis~\cite{maguire2017doing,newman2023human,huang2020ai} of the open-answer survey responses. One author reviewed the data and drafted an initial version of the codebook, while two other authors explored connections among these excerpts, gradually defining the themes. They are presented in bold as the initial title of thematic paragraphs in Section \ref{sec:results}.
The entire questionnaire form can be found in the Supplementary Material \url{https://ronfrancesca.github.io/PAGURI/}. {A block diagram summarizing the experimental procedure is shown in Figure~\ref{fig:study_procedure_diagram}.}
\begin{figure}[h]

\begin{tikzpicture}[
  node distance=1.2cm and 2.2cm,
  box/.style={rectangle, draw, rounded corners, minimum width=3.2cm, minimum height=1cm, align=center, font=\small},
  every edge/.style={draw, ->, thick},
  font=\small
]

\node[box, fill=myblue!40, text=white] (entry) {Participant Entry\\ (In-person or Online)};
\node[box, fill=myblue!50, text=white, below=of entry] (pre) {Preliminary Analysis\\ (Demographics + Gold-MSI questionnaires)};
\node[box, fill=myblue!60, below=of pre] (interact) {TTM Interaction\\ (PAGURI Interface)};
\node[box, fill=myblue!60, right=of interact, xshift=1.2cm] (eval) {Model Evaluation\\ (Per Iteration)};
\node[box, fill=myblue!80, below=of interact] (final) {Final Questionnaire\\ + Open Feedback};
\node[box, fill=myblue!100, below=of final] (analysis) {Thematic Analysis\\ (Open Coding)};

\draw[->] (entry) -- (pre);
\draw[->] (pre) -- (interact);
\draw[->] (interact) -- (eval);
\draw[->] (interact) -- (final);
\draw[->] (final) -- (analysis);


\end{tikzpicture}

    \caption{{Block diagram of the experimental procedure in the PAGURI user study. }}
    \label{fig:study_procedure_diagram}
\end{figure}
\section{Results}
\label{sec:results}


This section first presents the analytical data, followed by the key findings of the study. Due to space limitations, we present in this paper only the results we find most insightful. However, we have also included additional experiments and insights in the Supplementary Material, and we encourage the reader to explore them for a more comprehensive understanding of the results.

\subsection{Demographics Analysis of Participants}
\label{subsec:demographics}
A total of $24$ people participated in the study, with an average age of $26.9$ years \mbox{(SD = 5.01)}. $79.2\%$ identified as he/him, while $20.8\%$ identified as she/her.
To guarantee anonymity, each participant has been assigned an ID. 
95\% of the participants were of Italian nationality. 
In terms of current occupation, 62.5\% of the participants were master's degree students (mainly from the Master of Science in Music and Acoustic Engineering from Politecnico di Milano), 
4.2\% were bachelor's degree students, while the \mbox{remaining 33.3\%} were workers. Most participants are currently active in various musical endeavors: \mbox{$6$ of them} are part of a band or an orchestra, $6$ are music producers/mastering engineers or DJs, 2 of them are dance teachers, and one is a record label manager. 
62.5\% of the participants conducted the experiments in person, while 37.5\% participated remotely via the Zoom platform. We would like to emphasize that the experiment was made available both in person and online to reach the broadest possible pool of participants. 
It was the participant's decision whether to take the experiment online or in person.

\subsection{Participants' Musical Knowledge and AI Tool Experience: Demographics and Analysis}
\label{subsec:AIexperience}

\begin{figure}[h]
    \centering
    \includegraphics[width=\linewidth]{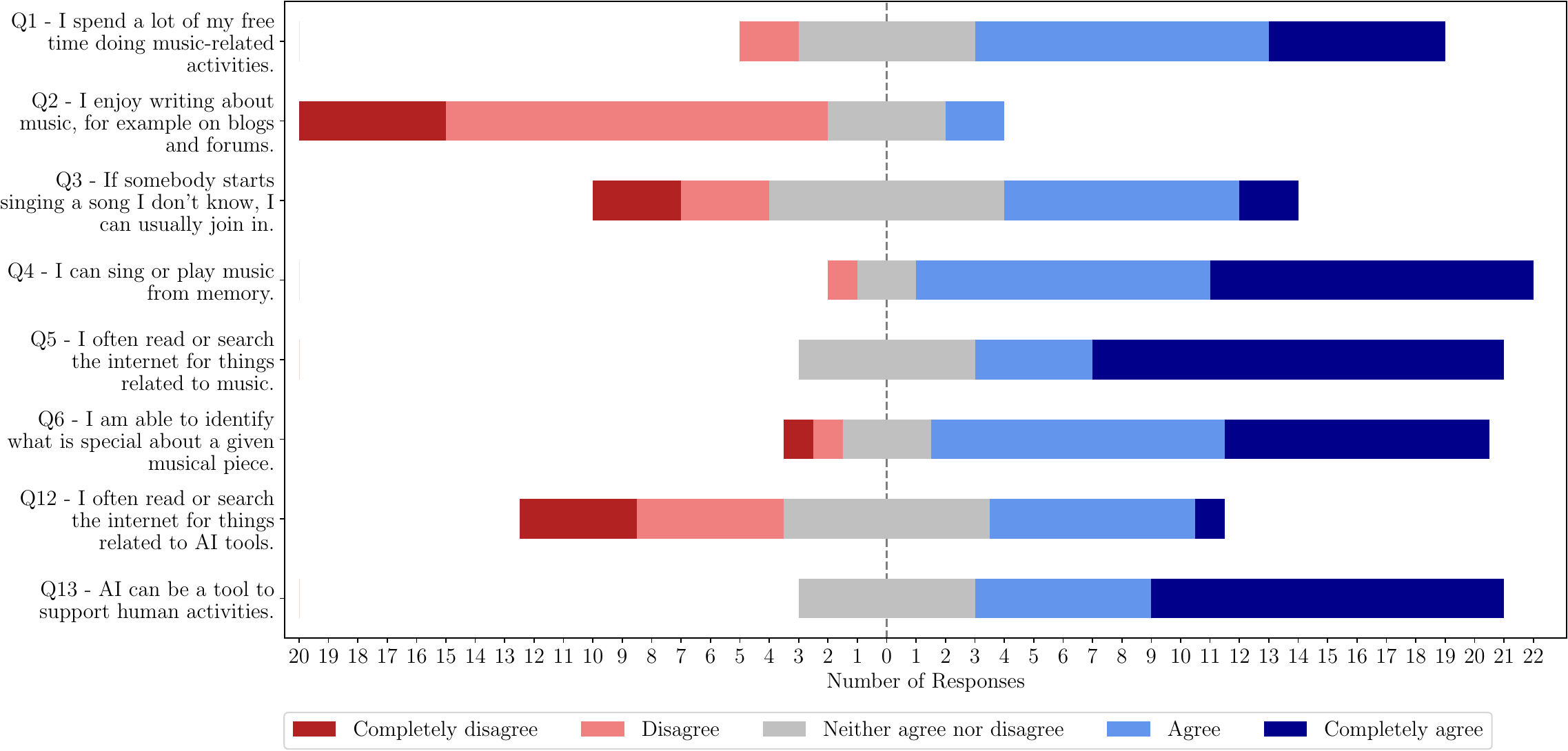}
    \caption{{Diverging} 
 bar chart showing the 
 \textit{musical knowledge and experiences with AI tools} \mbox{survey answers.}
    }
    \label{fig:MAISurvey}
\end{figure}

The first part of the experiment focused on understanding the relationship of participants with AI tools and music in general. To gather this insight, participants were asked to complete a questionnaire. 
Figure~\ref{fig:MAISurvey} presents a subset of questions rated on a five-point Likert scale, visualized using a diverging bar chart.
The chart illustrates the number of respondents for each answer, with positive responses on the right and negative responses on the left, centered around the midpoint. Neutral answers are split between positive and negative categories. The length of each bar represents the number of responses for each score, allowing easy comparison of the distribution. Each color indicates the group of responses corresponding to a specific score. Additional details on other responses are provided in the text.

Most participants demonstrated a strong interest in music, with $58.4\%$ listening to music for more than an hour daily. $74\%$ engaged in regular daily practice of an instrument for over years.
At the peak of their interest, more than $69\%$ of the participants practiced for at least an hour each day. Furthermore, $70\%$ of participants can play at least one instrument (such as guitar, piano, flute, drums, bass guitar, trombone, or ocarina) or are singers.

Regarding participants' familiarity with AI tools, $83.3\%$ reported being familiar with them, 
and $75\%$ 
could name specific examples, such as ChatGPT~\cite{wu2023brief}, \mbox{Copilot~\cite{bird2022taking}},\mbox{ Dall-E~\cite{ramesh2021zero}}, and Gemini~\cite{team2023gemini}. 
Some also mentioned TTM tools, such as Riffusion~\cite{forsgren2022riffusion} and MusicGen~\cite{agostinelli2023musiclm}.
When focusing specifically on TTM tools, $62.5\%$ were aware of their existence, though most could not name specific examples. Despite the widespread use of AI tools among participants, only $4.2\%$ reported actively using TTM models.

\subsection{Text-to-Music Interaction}
\label{subsec:TTM-int}


A total of 304 textual prompts were submitted to the TTM model (mean = 12.6 prompts per user, {Standard Deviation (SD)} = 4.56), leading to the generation of 1148 audio files (mean = 47.8 audio files per user; multiple audio files were generated per each prompt.) Additionally, there were 39 
personalizations made by participants 
(mean = 1.63 per user, SD = 0.75), meaning 39 customizations of one of the AudioLDM2 models (either the original AudioLDM2 or a previously personalized model) were done using the DreamBooth technique~\cite{plitsis2023investigating,ruiz2023dreambooth}. Of the 39 personalizations, 37 used the \textit{Fast 
}mode, while 2 used the \textit{Medium} mode. No one selected the \textit{Slow} personalization technique, likely due to the waiting time required. 
Different audio files were used as input to personalize the model, including electric guitar riffs, one-shot bass or synth lead samples, and 8--10 s tracks of musical beats or songs, among others. 
 
\textit{\textbf{Prompt ambiguity. 
}} When analyzing textual prompts, it was observed that many users referred to famous people, musical groups, everyday objects, or cartoon characters, 
expecting the generated audio tracks to be somehow related to them.
However, in most cases, the results did not align with their expectations. For instance, in the first iteration, P13 requested the model to generate music based on the prompt \textit{daft punk style song}, expecting electronic music in the style of the French electronic duo Daft Punk. However, the generated sound was rather reminiscent of punk rock. 
Similarly, in the $12$th consecutive generation of the same user, the prompt \textit{linkin park led to the creation of audio tracks featuring nature sounds or sounds associated with a natural park setting, rather than music resembling the rock style of Linkin Park}. Differently, audio tracks generated from textual prompts related to sound effects and soundtracks were highly appreciated by participants. P2 commented: \textit{``It is necessary to start from easy prompt to understand how to interact with the model, to then move to more complicated and sophisticated prompts''}. Most prompts referred to acoustic and electric guitar sound generation, drum beats, piano, and music styles. 
Additionally, many participants requested everyday sounds like cowbells or dog barking. Some users requested more abstract sounds, such as P22: \textit{underwater bossa nova classical guitar}, and P24: \textit{the sound of a drum in space}. Others asked for specific sounds, such as one of the prompts of P24: \textit{the sound of a 56k modem}.

\subsubsection*{Personalization of the TTM Model}
\label{subsec:personalization}

Most users, 22 out of 24 participants, personalized the TTM models by providing audio samples as input. They all started with the off-the-shelf pretrained \mbox{AudioLDM2 model. }

Overall, users highly valued the ability to tailor the model to their preferences. Below, we share some comments from the participants. If the code of the participant is followed \mbox{by $*$}, the sentence has been directly translated from Italian (this is valid for the whole paper). P4: \textit{``\textbf{It further improves the quality of the modeling the task of generating samples from a specific genre}. Even if this means that the quality in very different genres is decreased it's not a big deal, as we can still fall back to the original model or even fine-tune it again for a different task.''} P7: \textit{``It is good because \textbf{it can fit better your own music tastes}''}. P18: \textit{``The result is \textbf{more likeable} to \mbox{the user.}''} 

\textit{\textbf{Sound signature.}} Particular emphasis was placed on the ability of model personalization to shape a unique sound signature for the user.
P2: \textit{``It could be useful in order to \textbf{save your general signature sound} and use it to create \textbf{new sounds starting from them.}''} P13: \textit{``\textbf{It can provide signature sounds that only belongs to the artists.}''}

\textit{\textbf{Plagiarism.}} However, concerns regarding the copyright related to the personalization of the TTM model were also raised. P17 thinks that \textit{``the riff generated after customization is too similar to that of the input data, \textbf{at risk of plagiarism}.''} P5 commented, referring to the personalization, \textit{``it could also be a bit risky as it could essentially allow anyone to copy any style in very little time''.}
 
\textit{\textbf{Rhythmic and percussive personalization.}} Users were particularly impressed by the personalization ability of the models with respect to rhythms and percussive sounds. P3\textsuperscript{*}: \textit{``In terms of tone generation of the customization, \textbf{the drums are better reproduced.}''} P10\textsuperscript{*}: \textit{``With the basic model, the audio generated does not have a rhythm or cadence, while \textbf{with the customized model there is more rhythm and the tempo is better marked.}''} P14\textsuperscript{*} noticed that during the personalization, \textit{``the sound is very dissonant (bad even on a harmonic level), but \textbf{the percussive sounds are well reproduced.}''} P21\textsuperscript{*}: \textit{``It is very good at personalizing the rhythm.''}

\textit{\textbf{Multiple audio concepts conflict.}} Several users showed interest in repeatedly personalizing the same model by incorporating multiple text–audio concepts to test whether the model could handle multiple concepts simultaneously or if this would lead to an exclusive overlap of concepts. P2\textsuperscript{*} states that \textit{``Fine-tuning the model greatly increases the quality in the generation of one specific genre, but it also lowers the average quality when generating music from completely different genres.''} P5 claims that \textit{``The fine-tuning overlay \textbf{only works individually.}''} P13 expressed the desire to train the model with three different text–audio pairs (bass, drums, and synth lead sound) to generate a track containing all three sound classes. After completing the test, the participant stated that \textit{``the model can reproduce the desired sounds singular, \textbf{it fails to join all three of them effectively within a composition. I noted that there is always one of the three instruments that prevail over the others.''}} The same participant also noticed that when trying to combine multiple personalizations, there was an overlap of sounds, 
with the most recently customized audio dataset dominating. This issue is also known as catastrophic forgetting~\cite{plitsis2023investigating}.

\subsection{Consistency, Expectations, and Quality}
\label{subsec:expectations}

\begin{figure}[h]
    \centering
    \includegraphics[width=\linewidth]{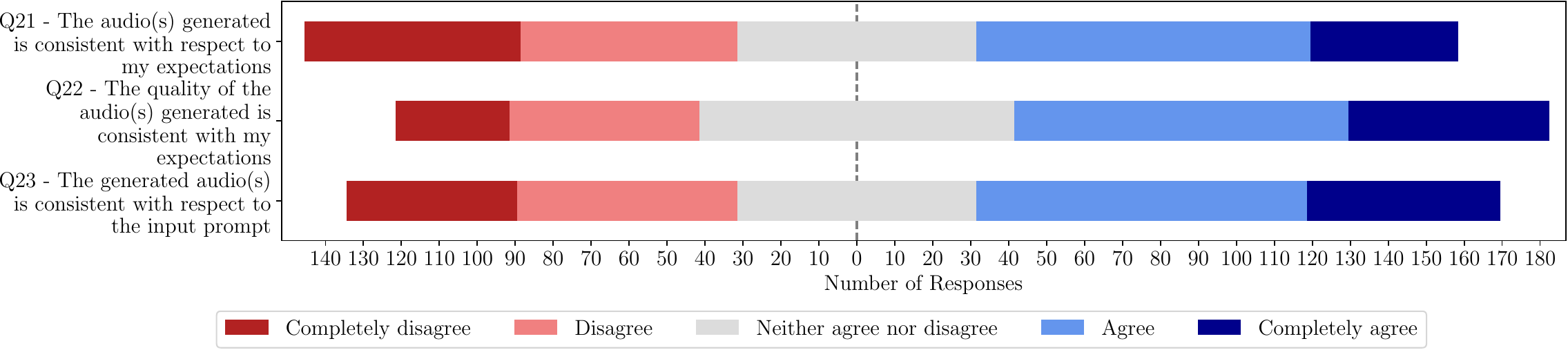}
\textls[-15]{    \caption{{Diverging} 
 bar chart showing the \textit{Model Evaluation Survey answers}. N.B. In this case, the value on the x-axis corresponds to the number of responses.  \label{fig:T2M_Survey}}}
   
\end{figure}

Each participant was asked to complete the survey \textit{Model Evaluation Survey} after each interaction with the model. Each audio(s) generation corresponds to one interaction with the model. Figure~\ref{fig:T2M_Survey} reports the answers for all the different interactions of all participants. 
The survey focused on understanding participants' perceptions related to three aspects: the consistency between the generated audio files and input prompts, the consistency between the generated audio samples and user expectations, and the overall quality of the audio(s) output concerning the user expectations. 
Only a limited number of the participants considered the overall generation to be entirely aligned with their expectations, both in terms of content and quality. The same holds for the consistency between audio and input prompts.  The majority of participants rated all three aspects between $3$ and $4$ (on a Likert scale from 1 to 5), suggesting that they perceived the generated audio to align reasonably well with their expectations and the input prompts provided. 
However, it is important to acknowledge that the subjective nature of the expectation scale presents a limitation in this study. We did not collect pre-generated user expectation data, but we intend to address this in future experimental designs.

\textit{\textbf{Audio generation expectation.}} Many participants were surprised by some of the audio files generated, with expectations ranging from the ability to sing a textual prompt (P5, P8, P11) to produce a spoken dialogue with understandable language (P11, P24) or a single instrument instead of a full musical beat (P7, P13). They also noted that the timbre of guitars and drums closely matched their expectations, resembling authentic instrument sounds. Many expected the model to adapt instruments, sounds, or melodies to a different musical style (i.e., performing style transfer), but this was not possible by design with the model/techniques considered. This indicates, in our view, that users should be informed and guided about the generation capabilities of TTM techniques.

\subsection{Example of Interaction with the Model}

As highlighted by the results discussed, drawing general conclusions is challenging, and it is crucial to consider the subjectivity of the interaction with TTMs. For these reasons, we report two examples of interactions of users with PAGURI by selecting them depending on their answer to question Q18 of the survey, namely \textit{What is your relationship with music?} We select P5, who answered \textit{DJ and/or music producer}, and P16, who instead answered \textit{I simply listen to music}. The rationale behind this choice is to loosely analyze the impact of musical knowledge on the subjective evaluation of the interaction with PAGURI. 
{Examples of interactions with the model for every user are available on the accompanying website {\url{https://ronfrancesca.github.io/PAGURI/interactions.html} (accessed on 25 August 2025). 

Figure~\ref{fig:example_interaction} reports the users' interactions with PAGURI by showing how the answers to the questions considered in Figure~\ref{fig:T2M_Survey} evolve during the experiment. Table~\ref{tab:example_interaction} shows the corresponding prompts used at each iteration by both considered participants. The blue bars indicate that the music generative model was not personalized during the corresponding iteration, while the red color indicates that the model has undergone the personalization procedure. As can be observed, both users begin by experimenting with the model and then proceed to personalize it with samples of choice before the third iteration. It is interesting how the two users interacted with the model. While P5, supposedly more 
musically experienced, experiments by applying effects to the personalized instrument, P16 seems to take the procedure less ``seriously'' by also experimenting with more incoherent prompts. 
\begin{figure}[h]

\subfloat[P5-Q21]{\label{fig:P5Q21}
\includegraphics[width=0.7\linewidth]{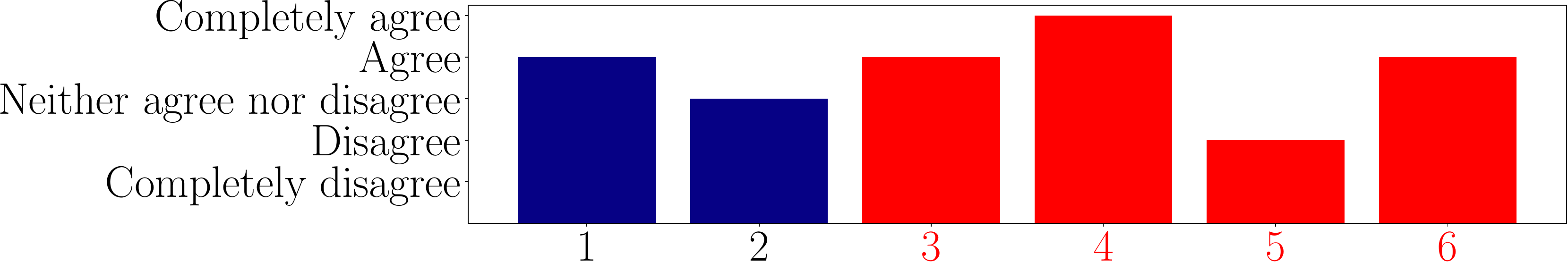}
}
\quad
\subfloat[P16-Q21]{\label{fig:P16Q21}
\centering
\includegraphics[width=0.7\linewidth]{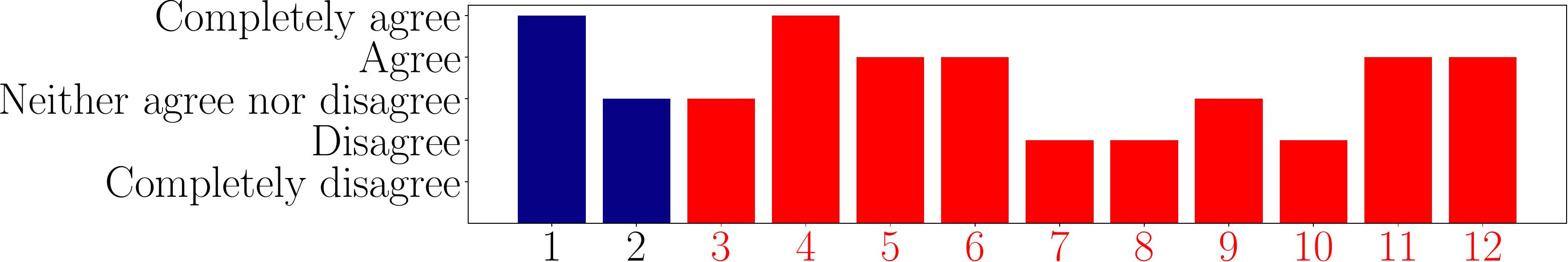}
}

\subfloat[P5-Q22]{\label{fig:P5Q22}
\includegraphics[width=0.7\columnwidth]{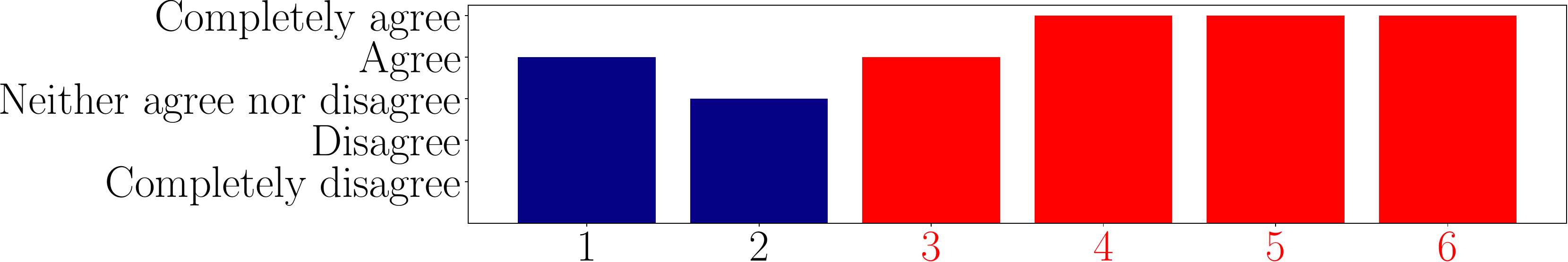}
}
\quad
\subfloat[P16-Q22]{\label{fig:P16Q22}
\includegraphics[width=0.7\columnwidth]{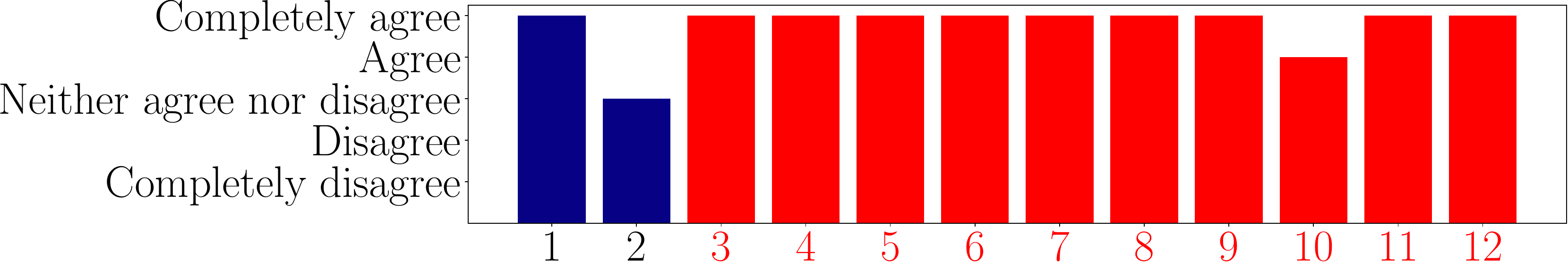}
}

\subfloat[P5-Q23]{\label{fig:P5Q23}
\includegraphics[width=0.7\columnwidth]{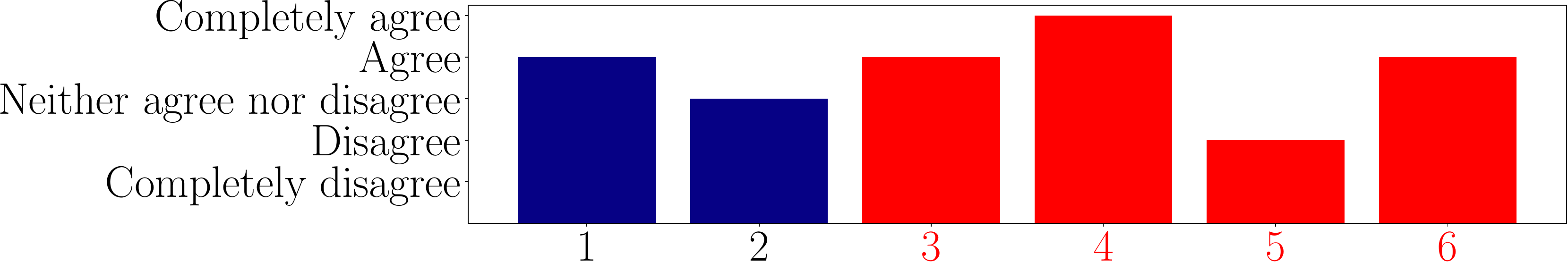}
}
\quad
\subfloat[P16-Q23]{\label{fig:P16Q23}
\includegraphics[width=0.7\columnwidth]{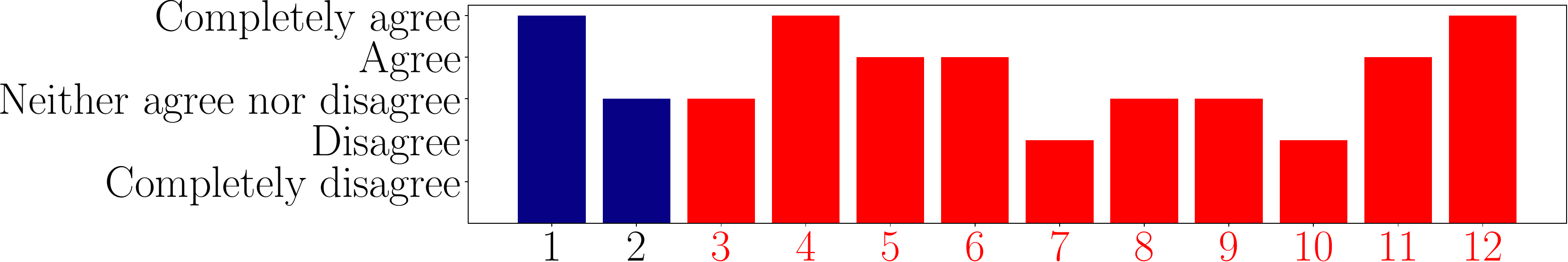}
}
\caption{{Answers of 
 P5 (\textbf{a},\textbf{c},\textbf{e}) and P16 (\textbf{b},\textbf{d},\textbf{f}) to the questions  Q21, Q22, and Q23 of the \textit{Model Evaluation Survey}, for each single interaction with the model. The red color indicates that at the specific iteration, the user personalized the model, the blue one, that personalization was not used. }}
\label{fig:example_interaction}
\end{figure}

\begin{table}[H]
\caption{{Prompts 
 used by P5 (\textbf{a}) and P16 (\textbf{b})} during the interaction with PAGURI. N.B. \textit{abcdef} and \textit{sas} are the pseudowords used to label and insert the new sounds during the personalization procedure of the TTM.}
\label{tab:example_interaction}

\begin{tabularx}{\textwidth}{cp{10cm}}
	\toprule
	\textbf{(a)}&\\\midrule
\textbf{Iteration} &\textbf{Prompt} \\\midrule

1 & A song about summer containing an electric guitar lead on top of a ukulele rhythm and a glockenspiel. The glockenspiel is off\\
\midrule
 2& Smooth jazz being played from the other room; the main line is played with a saxophone, backed by a Leslie organ\\\midrule
 3& An abcdef kalimba melody in a 90 BPM midtempo music drop\\
\midrule
 4&  An abcdef voice singing a Gregorian chant\\
\midrule
 5&  A kalimba with a harsh bitcrusher applied to it\\
\midrule
 6&  A abcdef kalimba with a bitcrusher applied to it\\\midrule
\textbf{(b)}&\\
\midrule
 \textbf{Iteration} &  \textbf{Prompt} \\
\midrule
 1&  A sound of an electric guitar\\
\midrule
 2&  A flute playing ``My Heart Will Go On'' by Celine Dion\\
\midrule
 3&  A ballad in the style of sas pirate metal\\
\midrule
 4&  A jazz music in the style of sas\\
\midrule
 5&  A doom metal music in the style of sas pirate metal\\
\midrule
 6&  A 90's disco hit in the style of sas pirate metal\\
\midrule
 7&  Vegeta singing a song in the style of sas pirate metal\\
\midrule
 8&  ``Epic sax guy'' in the style of sas pirate metal\\
\midrule
 9&  A dog barking in the style of sas pirate metal\\
\midrule
 10&  A dog barking in the style of pirate metal\\
\midrule
 11&  A rubber chicken singing a song in the style of sas pirate metal\\
\midrule
 12&  A sas pirate metal song, but with the lyrics made by a door bell ring\\
\midrule
\end{tabularx}

\end{table}
\subsection{Correlations Among Questionnaire Responses}
In this section, we report further analyses aimed at analyzing more granular relationships between the data. 

We computed Spearman’s correlations between all the questions and reported those with statistically significant p-values, obtained via a permutation test~\cite{ludbrook1998permutation}, to ensure the robustness of the results, given the relatively small sample size.

 In the case of the \textit{Model Evaluation Survey} shown in Figure~\ref{fig:T2M_Survey}, each user answered the same questions multiple times (i.e., after each model iteration). In this case, to compute correlations with the other questionnaires, we averaged the responses for each user. For tables~\ref{tab:aipercep},~\ref{tab:adoptionperception}, and~\ref{tab:expvsrea}, a row containing \texttt{"} indicates that the question is the same as in the previous row.

\begin{table}[H] 
  \caption{User experience and engagement (music and AI) correlations.    \label{tab:aipercep}}
 
 \begin{tabularx}{\textwidth}{p{4cm}|p{4cm}|c|c}
        \toprule
         \textbf{Question I} &  \textbf{Question II} &  \textbf{\textit{p}-Value} &  \textbf{Pearson’s R} \\
        \midrule
         Q6---I am able to identify what is special
about a given musical piece &  Q27---The audio generated by the personalized model is better than the audio generated by the based model &  0.0194 &   0.5 \\
\midrule
         Q9---At the peak of my interest, I practiced
\_\_\_\_\_ hours per day on my primary instrument &  Q24---I enjoyed the interaction with the text-
to-music generation system &  0.0166  &  0.5 \\
\midrule
         Q12---I often read or search the internet for
things related to AI tools &  Q24---I enjoyed the interaction with the text-
to-music generation system &  0.0292  &  0.46 \\
\midrule
        " 
 &  Q25---The data generated by the model are
consistent with respect to the desired audio
file provided by the user &  0.0254 &  0.48 \\

\midrule
        " &  Q27---The audio generated by the personalized model is better than the audio generated by the based model &  0.0428 &  0.44 \\
        \bottomrule
    \end{tabularx}
\end{table}
\unskip
\begin{table}[H] 
\caption{AI perception-related correlations.}
    \label{tab:adoptionperception}
  
 \begin{tabularx}{\textwidth}{p{4cm}|p{4cm}|c|c}
        \toprule
         \textbf{Question I} &  \textbf{Question II} &  \textbf{\textit{p}-Value} &  \textbf{Spearman’s \boldmath{$\rho$}} \\
        \midrule
         Q13---AI can be a tool to support human
activities &  Q24---Enjoyment of T2M system &  0.0019  &  0.61 \\
\midrule
        " &  Q25---Model output consistency with desired audio &  0.0109  &  0.49\\
\midrule
       " &  Q29---T2M models support music creation &  0.0037  &  0.55 \\
\midrule
        " &  Q30---Willingness to use the system again &  0.0055 &  0.53 \\
        \bottomrule
    \end{tabularx}
\end{table}
\unskip
\begin{table}[H]
 \caption{Correlations concerning expectations vs. reality.}
    \label{tab:expvsrea}
   \begin{tabularx}{\textwidth}{p{4cm}|p{4cm}|c|c}
        \toprule
         \textbf{Question I} &  \textbf{Question II} &  \textbf{\textit{p}-Value} &  \textbf{Spearman's \boldmath{$\rho$}} \\
        \midrule        
         Q21---The audio(s) generated is consistent with respect to my expectations &  Q26---The waiting time of the fine-tuning of
the model is proportionate with the quality
of the generated audio &  0.0458 &  0.43 \\
\midrule
        " &  Q28---There is consistency between input
prompt and audio(s) generated &  0.0006 &  0.66 \\
\midrule
         Q22---The quality of the audio(s) generated is consistent with my expectations &  Q24---I enjoyed the interaction with the text-to-music generation system &   0.0092 &  0.53 \\
         \midrule
         " &  Q25---The data generated by the model are consistent with respect to the desired audio file provided by the user &  0.0048 &  0.58 \\
        \midrule
        " &  Q26---The waiting time of the fine-tuning of the model is proportionate with the quality of the generated audio &  0.015 &  0.52 \\
        \midrule
        " &  Q28---There is consistency between input prompt and audio(s) generated &  0.0032 &  0.58 \\
        \midrule
        " &  Q29---The use of text-to-music models can support musicians in musical creation endeavours &  0.0426 &  0.42 \\
        \midrule
        " &  Q30---I would use this system again &  0.007 &  0.55 \\
        \midrule
         Q23---The generated audio(s) is consistent with respect to the input prompt &   Q28---Generated audio consistency with prompt vs. consistency of generated audio &  0.0028 &  0.6 \\
       
        \bottomrule
    \end{tabularx}
\end{table}

The correlations presented in Table \ref{tab:aipercep} highlight key insights into user engagement and perceptions of the TTM generation system. Significant positive correlations suggest that higher engagement with music and AI-related topics (Q9, Q12) is associated with greater enjoyment and perceived quality of the generated audio. The correlation between Q9 and Q24 reflects a strong relationship between musical practice and the enjoyment of the output of the model. 
These results highlight the connections between musical engagement, AI-related interests, and users' positive evaluations of the text-to-music system.

\subsubsection{AI Perception and Adoption}
\label{subsec:perceptionadoption}

Table~\ref{tab:adoptionperception} presents the correlations related to AI perception and its potential adoption. Significant positive correlations are observed between Q13 and Q24, Q25, Q29, and Q30. These findings suggest that users who perceive AI as a useful tool for supporting human activities also tend to have higher enjoyment of the system, appreciate the consistency of model outputs, and recognize the potential of T2M models for music creation. Additionally, users who see AI positively are more likely to express a willingness to use the system again.

\subsubsection{Expectations vs. Reality (Consistency and Quality)}
\label{subsec:expvsrea}

Table \ref{tab:expvsrea} illustrates a series of significant correlations between user expectations and their experience with the text-to-music generation system, providing insights into how certain aspects of the system’s performance align with user perceptions. There is a clear trend suggesting that users' expectations regarding audio consistency and its quality are highly correlated with their overall satisfaction and willingness to engage with the system again. The correlation between Q21 and Q28, which links expectations of audio consistency to the alignment between input prompts and generated audio, further underscores the importance of maintaining a coherent and predictable relationship between user input and system output. The data also suggest that fine-tuning aspects of the model, such as the waiting time for improvements (Q26), significantly affects how users perceive the quality of the generated audio. There is a noticeable positive correlation between Q22 (audio quality consistency) and Q24 (user enjoyment), suggesting that the perceived quality of the system's output enhances user satisfaction. These correlations point to a pattern where consistency, quality, and perceived utility are intertwined, with each factor contributing to the users’ enjoyment and their propensity to continue using the system. This suggests that refining the system’s ability to meet user expectations, particularly in terms of audio consistency and fine-tuning responsiveness, may significantly enhance both user satisfaction and the perceived value of the tool in creative processes.

\subsection{Integration of TTM Models in the Creative Process}
\label{subsec:integration}

In Figure~\ref{fig:final_survey}, we present responses to Likert scale questions related to the final questionnaire. Q24 shows unanimous agreement on enjoying the
interaction with the system, reflecting strong user engagement. However, Q28 highlights a key
challenge for current TTM models, as no participants completely agreed on the consistency
between the input prompt and the audio generated. Interestingly, as Q26 shows, none of the
participants completely disagree that the fine-tuning waiting time is proportionate to the output
quality. In Q31, we asked if they would include this workflow in their music creative process and how. $33.5\%$ of participants selected \textit{I would only use it for inspiration purpose}, while $37.5\%$ chose \textit{I would take the audio generated and modify it in post-generation}. Other participants shared different comments. P17\textsuperscript{*} asserted: \textit{``At the moment, no, because the program does not process music according to my specifications''}, suggesting the need for higher controllability. P10\textsuperscript{*} commented, \textit{``I would use it for specific exercises on rhythm and improvisation based on the generated audio''}, which is a type of application that was not considered when we designed the questionnaire. 
\begin{figure}[h]
    \centering
    \includegraphics[width=.9\linewidth]{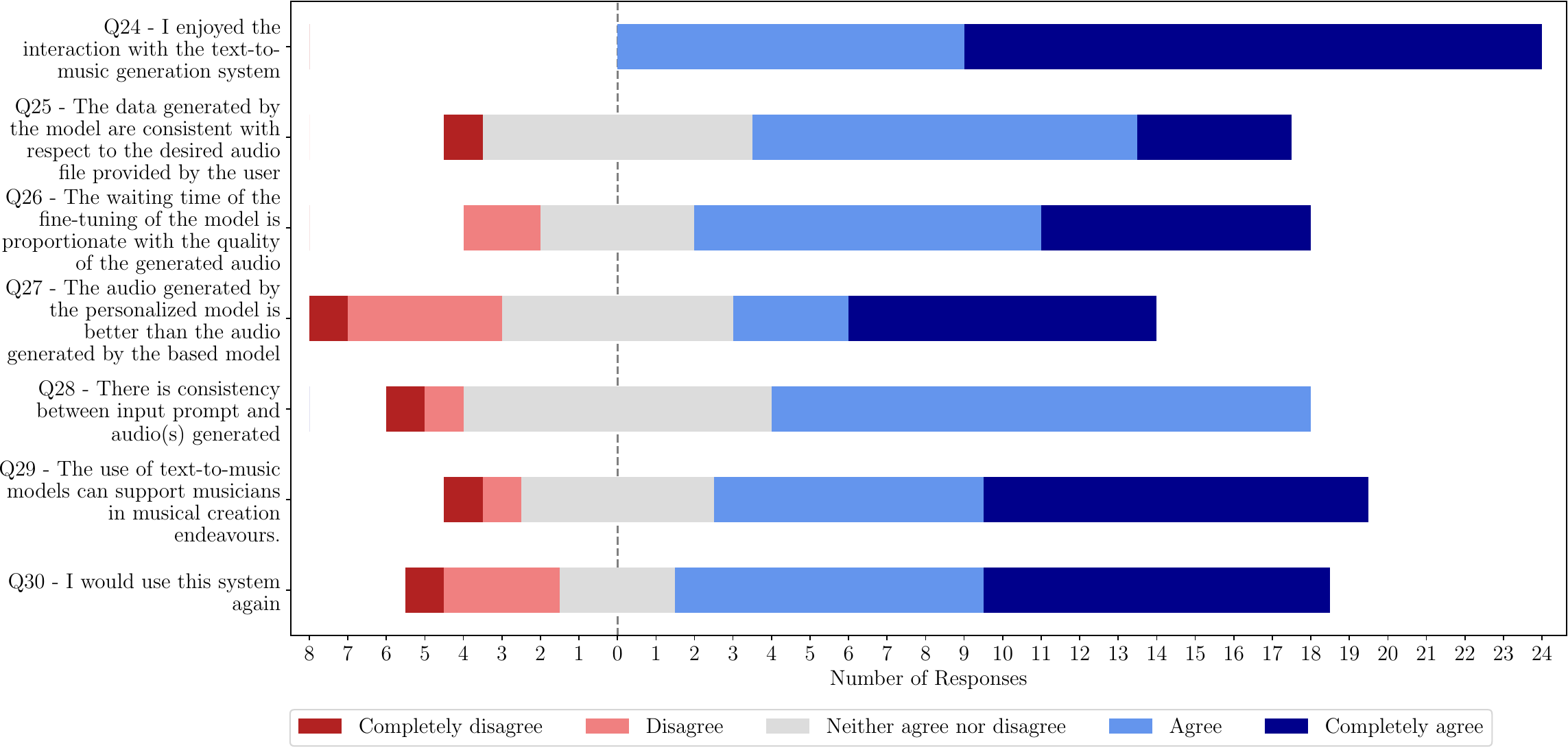}
    \caption{Diverging bar chart showing the \textit{Final Satisfaction Survey answers}. 
    Q25, Q26, and Q27 report results not considering users who did not personalize the model.}
    \label{fig:final_survey}
\end{figure}
\textit{\textbf{TTM possible applications.}} In the same questionnaire, we asked about the contexts in which users would use the audio generated by the model, both in the case of original and personalized models. Most participants gave expected answers related to the use of samples/loops/beats for music production, for sound design, or as an inspiration tool. Others provided original suggestions, probably related to their personal work and creative context, such as the use for Foley sound generation, or P13\textsuperscript{*}: \textit{``during dance lectures''}, and P14\textsuperscript{*}: \textit{``I would use it to build musical tracks for dance choreography''}. The latter two are particularly interesting, suggesting that when considering the possible applications of TTM models, 
it is necessary to consider the user's background, who may have a final application in mind that the model designers did not originally envision.
Some participants, referring to the personalized model, suggested that \textit{``The personalized model \textbf{must adhere more closely to the specifications to be used}''} (P17\textsuperscript{*}) or \textit{``I would not use it because \textbf{it did not meet my expectations}''} (P10\textsuperscript{*}). This suggests that the model's adherence to the desired objectives is important to users.

P14 suggested, \textit{``I would use it to write music from a particular sub-genre''}. This is interesting because while most models can generate music belonging to mainstream genres, some users might require the ability to generate music from specific sub-genres, which are often underrepresented in music datasets typically used to train these models. A feature like personalization would allow the inclusion of sub-genres into the models, enabling the fine-tuning of the model with sub-genres it has not encountered before.

Key insights focused on the role of TTM models in the democratization of AI tools for music production.  P1: \textit{``I believe that integration of artificial intelligence can give a boost to music production, providing \textbf{more opportunities for young artists.}''} 

\section{Discussion}
\label{sec:discussion}

In this section, we present a summary of the research findings from the proposed study, providing key insights into users' perceptions of TTM models and their impact on the creative process, {and also highlighting the issues and challenges of integrating such models into real-world artistic practice}.

\textbf{\textit{Personalization and reference audio play an important role in TTM models.}} As highlighted in Section~\ref{subsec:TTM-int}, user personalization is a key factor in TTM models. Although users often have a specific audio reference in mind, as shown by using famous artists or music bands to guide the generation process, relying solely on text prompts is limiting and does not always produce the desired audio results. In contrast, providing an audio clip that reflects a specific style or sound can be a more effective guide for the model, simplifying the process and yielding more accurate results. Thus, incorporating reference audio should be a standard feature in all TTM models, as demonstrated by recent advancements~\cite{nistal2024diff,demerle2024combining,lin2023content}, which acknowledge the inherent limitations of text-only control in music. Additionally, personalization techniques should be recognized as essential for enhancing user experience and creative control, and they should be integrated into TTM models. However, these techniques often face challenges, such as catastrophic forgetting~\cite{plitsis2023investigating}, which can impair model performance over time. Addressing this issue is crucial to ensuring that TTM models retain learned features while adapting to new user inputs~\cite{bafghi2025fine,feng2025zeroflow}.

\textbf{\textit{High-quality generated samples are not all that matters.}} Sometimes, they are expected by users, especially in music production, but there are contexts where they are not essential.  In music creation, for instance, decent audio samples from TTM models can often suffice to stimulate creativity and be directly integrated into musicians' workflows. 
Even if opinions regarding the consistency of audio quality in relation to expectations vary widely, both positively and negatively (cf. answers to Q22 in Figure~\ref{fig:T2M_Survey}), participants stated that they would use the generated sounds directly, as they are, for inspiration, beat-making, music production, and even for crafting tracks for dance practice. This suggests that users who use TTM models to generate sounds for inspirational purposes may not need high-quality sound-generated audio. Instead, they might prioritize the creativity and novelty of the generated sounds over quality. Future TTM models should focus on creativity, diversity, and customization, balancing sound quality for different creative purposes. Furthermore, incorporating subgenre-specific datasets alongside traditional ones and increasing diversity should be a priority, as also reported in~\cite{chen2024applications}.

\textbf{\textit{Context and subjective perception are important when evaluating TTM models.}}
Analyzing the perception of TTM systems based on different contexts and with participants with different backgrounds made clear it how challenging it is to objectively evaluate TTM models. This finding is consistent with the results presented in~\cite{vinay2022evaluating} regarding the limitations of objective metrics in evaluating generative audio systems. Metrics commonly used to evaluate generative models, such as the Fr{\'e}chet Audio Distance~\cite{kilgour2019frechet}, while informative, often fail to capture the nuances of user interactions and preferences. Assessing usability requires a thorough understanding of how these systems are used and perceived by end-users, as also highlighted in~\cite{chu2022empirical} in the context of automatic symbolic music generation models. 
Determining the usefulness of TTM model outputs is not straightforward when relying only on objective metrics, suggesting the need for novel and complementary subjective evaluative approaches that also consider usability, user-centered experience, and contextual relevance. To assess the value and impact of TTM models, it is essential to incorporate subjective evaluations, including feedback from artists, not only in terms of quality through listening tests but also to ensure that the outputs and the control offered align with their creative needs and are truly effective in real-world contexts.

\textbf{\textit{TTM models contribute to democratizing the music creation process but raise concerns regarding copyright protection.}} Some participants emphasized how the model and the personalization technique explored in this study empower individuals to create their unique sound signatures, enabling a user-centered audio generation. However, concerns arise about control and copyright protection. As also pointed out by some participants, with personalization techniques, it becomes easy to incorporate others' signature sounds into the generative model, making them vulnerable to unauthorized use. While music AI-based systems democratize music creation, making it accessible to a wider user base, there is a crucial need to address ethical guidelines and copyright laws regarding training datasets and the use of generated sounds~\cite{ren2024copyright,franceschelli2022copyright}. 

\textbf{\textit{Controllability and editability are at the forefront of requested features.}}
One limitation of the model considered in this study is the lack of editability, preventing users from actively influencing the direction of the audio generation process based on their choices. This issue is not unique to this model, as other systems also fail to provide sufficient control or editability over the generated audio output~\cite{copet2024simple,huang2023make}. Participants expressed a desire for greater control over the generation process, such as the ability to modify specific sections of the generated music track, not only limited to style or genre transfer~\cite{zhang2024musicmagus}. Therefore, it is crucial that future TTM models include user-friendly interfaces that enable users to adjust and fine-tune specific parameters to guide the creation of the desired audio. Moreover, the graphical interfaces cannot be the same for every user but need to be contextualized according to the final users and their final tasks. In fact, some of these features have been integrated into the just-released models (at the time of writing), such as Stable Audio 2.0 (from Stability AI) and Project Music GenAI Control (from Adobe). 
An alternative and promising direction is to introduce control mechanisms based on symbolic representations of music, such as melody, rhythm, or harmony, making them essential features alongside text-based prompts. Recent studies have begun exploring this approach, reinforcing the necessity of providing more direct control over musical attributes rather than relying solely on text prompts~\cite{wu2024music,demerle2024combining}.

\subsection*{User Study Limitations}
As with any user study, PAGURI also suffers from limitations, some due to circumstances 
some due to inherent limitations of text-to-music and generative models in general.

We acknowledge that the diversity of nationality and education level of the pool of participants is limited. However, we believe that their interest and knowledge of music and AI-related tools make the results, their comments, and suggestions interesting to the wider research community. Even though we tried to enlarge the number of participants by publishing the advert on several relevant mailing lists and by providing the opportunity to participate online, the time-consuming nature of the experiment made it hard to reach a wider public.

The limitations in terms of diversity can also be translated to the TTM models considered. Indeed, the music generated by the AudioLDM2 model, as well as similar models~\cite{copet2024simple,kreuk2022audiogen}, frequently reflects mainstream Western music, which aligns with the cultural context in which these models were trained and the diversity of the data used during training. These datasets, even if mostly undisclosed, are often based on Western music genres and, consequently, do not generate music from underrepresented world cultures. 

\section{{Conclusions}}
\label{sec:concl}
This study aims to examine how users interact with text-to-music generative models, specifically focusing on whether they would incorporate these tools into their creative workflows. Additionally, it investigates whether multimodal input could enhance both the model's performance and the users' creativity. Through the interface, study participants were able to generate audio from textual prompts using a selected TTM model and/or personalize the model to generate specific sounds based on their preferences. Throughout the study, we gathered feedback through questionnaires, asking targeted questions about their interaction with the interface and their overall perception of TTM models.
The findings highlight that TTM models offer diverse application opportunities beyond music creation, with a strong emphasis on the need for greater personalization and user control. Participants expressed enthusiasm for the potential of these tools but also raised concerns about plagiarism and the unauthorized use of generated sounds.  Additionally, there is a clear need to increase diversity in audio generation, moving beyond a Western-centric focus, and to provide users with greater control to better express their creative intentions. Text-based prompts alone are limiting, highlighting the importance of alternative input methods. Furthermore, subjective evaluation should go beyond traditional listening tests assessing quality, as usability and creative relevance are equally critical factors.

\subsection*{Future Work}
While the results of the PAGURI user study provide some important insights related to the interaction between users and TTMs, we believe that they also lay the ground for future developments of the procedure.
Future works aim to include these results in TTM generative models and develop a better interface to allow further interaction and control for the final users. {Also, we plan on allowing participants to test a number of diverse TTM models in order to be able to analyze how their usability varies depending on the controls allowed and on the type of music on which the models were trained.} As an example, it could be interesting to allow users to have the possibility of using multimodal inputs to the models. 
Furthermore, we also plan on performing more complex longitudinal studies with an additionally developed version of PAGURI in order to study how TTMs could be used over a longer time span, thus effectively simulating how end users could leverage these models for music production practices. Overall, results suggest that text-to-music models can serve as valuable tools for creative workflows and educational settings, provided that several challenges such as controllability, personalization, and workflow integration are addressed and deserve to be considered in such scenarios in future research.

\small \textbf{Author declare that they do not have any competing interests.}
%
%
%
\bibliographystyle{splncs04}
\bibliography{biblio}
%




\end{document}